\documentclass[journal=aamick,manuscript=article]{achemso}
\usepackage{amsmath}
\usepackage{amssymb}
\usepackage{graphicx}
\usepackage{dcolumn}
\usepackage{bm}
\newcommand{\vect}[1]{\boldsymbol{#1}}
\title{Prediction of transport property via machine learning molecular movements}

\author{Ikki Yasuda}
\affiliation{Department of Mechanical Engineering, Keio University, Yokohama, Kanagawa 223-8522, Japan}
\author{Yusei Kobayashi}   
\affiliation{Department of Mechanical Engineering, Keio University, Yokohama, Kanagawa 223-8522, Japan}
\author{Katsuhiro Endo}   
\affiliation{Department of Mechanical Engineering, Keio University, Yokohama, Kanagawa 223-8522, Japan}
\author{Yoshihiro Hayakawa}   
\affiliation{Department of General Engineering, National Institute of Technology, Sendai College}
\alsoaffiliation{Deceased January 11. 2020}
\author{Kazuhiko Fujiwara}   
\affiliation{Department of General Engineering, National Institute of Technology, Sendai College}
\author{Kuniaki Yajima}   
\affiliation{Department of General Engineering, National Institute of Technology, Sendai College}
\author{Noriyoshi Arai}   
\affiliation{Department of Mechanical Engineering, Keio University, Yokohama, Kanagawa 223-8522, Japan}
\author{Kenji Yasuoka}   
\email{yasuoka@mech.keio.ac.jp}
\affiliation{Department of Mechanical Engineering, Keio University, Yokohama, Kanagawa 223-8522, Japan}

\begin{document}
\begin{abstract}
Molecular dynamics (MD) simulations are increasingly being combined with machine learning (ML) to predict material properties. The molecular configurations obtained from MD are represented by multiple features, such as thermodynamic properties, and are used as the ML input. However, to accurately find the input--output patterns, ML requires a sufficiently sized dataset that depends on the complexity of the ML model. Generating such a large dataset from MD simulations is not ideal because of their high computation cost. In this study, we present a simple supervised ML method to predict the transport properties of materials. To simplify the model, an unsupervised ML method obtains an efficient representation of molecular movements. This method was applied to predict the viscosity of lubricant molecules in confinement with shear flow. Furthermore, simplicity facilitates the interpretation of the model to understand the molecular mechanics of viscosity. We revealed two types of molecular mechanisms that contribute to low viscosity. 
\end{abstract}

\section{Introduction}
Simulation and data-driven methods are two major approaches in material informatics, and the two approaches are increasingly combined to achieve an efficient material design in silico \cite{ramprasad2017machine}. In particular, the combination of molecular dynamics (MD) simulation and machine learning (ML) is being increasingly used for various applications \cite{noe2020machine}. The MD simulations generate molecular configurations of materials, which provide their structure and dynamics, and enables the calculation of various material properties, such as thermodynamic and mechanical properties. Although traditional approaches primarily employ physics-based analysis, recent ML methods have proposed data-driven approaches, thereby expanding the application of molecular data. For example, ML-based force fields \cite{behler2016perspective,deringer2019machine,unke2021machine} have enhanced MD simulations with an accuracy similar to that of quantum mechanics-based force fields, namely for molten salts \cite{nguyen2021actinide,liang2021machine, rodriguez2021thermodynamic}. In another application, ML methods have predicted chemical and mechanical properties of various chemical compounds, polymeric membranes, and silicon surfaces using MD trajectories via the combinations of other properties such as potential energies \cite{riniker2017molecular,kirch2020brine,ritt2022machine} and neighboring environments of atoms \cite{bartok2017machine}. Machine learning have also been used to extract transitions in metastable states from the complex and vast amount of MD trajectories in biomolecules \cite{mardt2018vampnets,lohr2021kinetic} and the solid-liquid interface of silicon \cite{xie2019graph}. Recently, lubricant molecules designed by ML have been evaluated using MD, and the results have been fed back to build an autonomous learning cycle \cite{kajita2020autonomous}. 

Material informatics using ML significantly depends on the availability of datasets. The limitation of available datasets decreases the accuracy of its prediction accuracy \cite{faber2016machine,schmidt2017predicting}. Considering that a sufficient dataset is not always available for the material properties of interest despite the growth of material platforms \cite{himanen2019data,allers2021artificial}, ML approaches that efficiently use datasets are in high demand \cite{zhang2018strategy}. In the point of dataset availability, ML combined with MD might not suffer from the lack of the datasets because a diverse amount of MD data can be generated with slight changes in simulation conditions \cite{wang2019mechanical}. That said, MD calculations are still computationally expensive, and sampling thousands of systems is not realistic which might be necessary to train complicated ML models \cite{perez2018simulations,rahman2021machine}. Thus, MD--ML methods using simple models are expected to be used in diverse applications when the dataset availability is limited.

Transport properties, which describe the movements of mass and heat, have critical roles in materials of various products, thus being widely investigated using MD and ML methods \cite{chakraborty2020quenching,liu2021high}. For example, the diffusion coefficient have been predicted for gases molecules \cite{ritt2022machine}, ions \cite{sendek2018machine,wang2022all} and other chemical compounds \cite{allers2021using} to discover and design better materials for artificial membrane and batteries. Similarly, the viscosity measures the ability of lubricants to reduce friction and wear in machinery. Considering that friction and wear are related to approximately a quarter of the global energy consumption, improvements in lubricants can significantly reduce the amount of wasted energy \cite{holmberg2017influence}. To measure viscosity and understand the molecular mechanism of friction reduction by lubricant molecules, non-equilibrium MD (NEMD) simulations of shear flow have been performed \cite{ewen2018advances,ewen2021contributions}. In particular, the NEMD simulations for confined lubricant films can simulate nanoscale contact friction during rolling and sliding of materials \cite{urbakh2004nonlinear}, wherein lubricant molecules were confined in thin spaces equivalent to a few molecular layers. Lubricants under shear stress in confinement exhibit shear thinning, i.e. their viscosity decreases with increasing shear rate. \cite{spikes2014history}. 

In this study, we present an ML method to predict the transport property of lubricant molecules using relatively scare training datasets obtained from MD and a simple supervised ML model. The simplicity of the model is achieved by an efficient representation of the molecular movements using an unsupervised ML method developed by Endo et. al. \cite{endo2019detection}. First, in the unsupervised ML method, we compared the molecular movements in pairwise systems with differing molecular species and simulation conditions. Based on the relative differences in molecular movements in the pairwise systems, a global relation of all the systems was determined in a low-dimensional space. Second, the low-dimensional representations of the systems were used as the input for the supervised ML to predict the material properties of interest. Furthermore, the simplicity of the prediction model enables the interpretation of the predicted property, thereby providing direct links between the molecular mechanism and the predicted property. We verified our approach using four types of lubricants molecules in confined systems with four different shear velocities for each type. We demonstrate that our method predicts the correct shear viscosity regardless of the molecular species. The interpretation of the predicted viscosity reveals the detailed molecular mechanics of shear thinning.    

\section{Methodology}
\subsection{Local Dynamics Ensemble: Representation of Molecular Movements}
In this study, local dynamics ensemble \cite{endo2019detection} (LDE) was used to represent the molecular movements obtained from the MD simulations, which was then analyzed using a series of ML methods (Fig. 1). The LDE is an ensemble of short-term trajectories $\vect{x}$,
\begin{equation}
    \vect{x} = \left[\vect{r}(t_0+\Delta)-\vect{r}(t_0), ..., \vect{r}(t_0+\delta)-\vect{r}(t_0), \vect{0}   \right]
    \label{eq:x}
\end{equation}
where $\vect{r}$ is the position of the particles of interest, $t_0$ is the time step in MD simulation, $\Delta$ is the LDE time, and $\delta$ is the time interval of MD output. In this study, the particles of interest were the stream-wise position of the geometrical molecule center. The LDE time and $\delta$ was 640 ps and 10 ps, respectively (Fig. 1a). As the transport properties are dictated by the movement of molecules, we have assumed that Equation (\ref{eq:x}) is appropriate for the prediction of transport properties. The MD details and the LDE selection methodology are described next. 

The MD data were obtained from the NEMD simulations of confined lubricant films by Kobayashi et. al \cite{kobayashi2021}, and included 16 different systems in four molecular types (Hexadecane, Octamethylcyclotetrasiloxane (OMCTS), Polyalphaolefin (PAO), and Pristane, see Fig. \ref{fig:mols}) and four different wall sliding velocities $V$ ($10^{-3}$,$10^{-2}$, $10^{-1}$, $1$ m/s). The lubricant molecules were confined in atomically smooth walls made of silica, and the lubricant thickness was limited to two molecular layers. The molecules were modeled using united atom models, and the OMCTS was assumed to be a rigid body. The number of molecules ranged from 77 to 210 depending on the molecular species and the sampling time of MD was more than 1.4 ns in the steady state, as determined from the calculated viscosity.

The LDE was employed for the following reasons. First, the four lubricant species are substantially different in terms of their structure, which makes it challenging to compare the dynamics at the atomic level. To overcome the differences, The LDE treats each molecule in an identical format, i.e. through its geometrical center. Furthermore, we assumed that the molecular translation and rotation, and not the deformation are important for transport properties because the former is dominant in the long-term molecular movements. As shown in Fig. \ref{fig:disp} and Fig. S1, the major molecular characteristics of the all-particle models at 640 ps in the streamwise direction could be captured using just the geometric center. Second, a temporal trend of the dynamics can be represented using the LDE. This not possible using conventional diffusion analysis such as root mean square displacement. Third, the LDE can represent the molecular movements using moderately sized data. This makes it possible to trest molecular movements in the limited sized nodes of deep neural networks (DNNs).

\subsection{Deep Learning to Measure Differences in LDEs}
The difference in the LDEs between a pair of systems $i$ and $j$ was measured based on the Wasserstein distances \cite{villani2009optimal,arjovsky2017wasserstein} (Fig. 1b), 
\begin{equation}
    W_{ij}= \sup_{||f||_{L \leq 1} } \mathbb{E}_{\vect{x} \sim \vect{y_i} }\left[ f_{ij}(\vect{x})\right] - \mathbb{E}_{\vect{x} \sim \vect{y_j}}  \left[f_{ij}(\vect{x})\right] \label{eq:wd1}
\end{equation}
where supremum is taken over all the 1-Lipschitz function $f_{ij}$, $\vect{x}$ is the short-term trajectory of the LDE, and $\vect{y_i}$ is the probability distribution of the LDE of system $i$. The function $f_{ij}$ can be approximated by a DNN, and the trained network $f_{ij}^*$ yields,  
\begin{equation}
W_{ij}= \mathbb{E}_{\vect{x} \sim \vect{y_i} }\left[
f^*(\vect{x})\right] - \mathbb{E}_{\vect{x} \sim \vect{y_j}} \left[f^*(\vect{x})\right]. \label{eq:wd2}
\end{equation}
The architecture of the DNN was adapted from Endo et al. \cite{endo2019detection}. Briefly, the DNN for the function $f_{ij}$ consists of a flattening layer as the input layer, three affine layers with 1024 output nodes and a rectified linear unit (ReLU) activation function as the latent layer, and an output layer with one node. The output layer had neither a bias term nor activation function. The loss function with a gradient penalty was used \cite{gulrajani2017improved}. The model parameters were updated using the Adam optimizer \cite{kingma2014adam} until the moving averages of the output became stable. During the training of the DNNs, the short-term trajectories were randomly selected from all the molecules and from time steps.  

The Wasserstein distance was calculated for all pairs of systems and a distance matrix in the ($N$, $N$) dimension was obtained, where $N$ is the number of systems. Because the matrix shows the relative differences between pairwise systems, the global relation of the $N$ systems can also be estimated from the matrix (Fig. 1c). Therefore, the matrix was converted into $N$ vectors in $n$ dimensional space such that the Euclidean distances best represent the Wasserstein distances,
\newcommand{\argmin}{\mathop{\rm arg~min}\limits}
\begin{equation}
    \vect{p}_0,\vect{p}_1,...,\vect{p}_n= \argmin_{\vect{p}_0,\vect{p}_1,...,\vect{p}_n} \sum_{i,j}(W_{i,j}-||\vect{p}_i-\vect{p}_j||)^2. \label{eq:emb}
\end{equation}
The embedded vectors $\vect{p}_0,\vect{p}_1,...,\vect{p}_n$ were optimized using simulated annealing to explore the global minimum. Subsequently, the gradient descent method was used to for fast convergence. After that, a principal component analysis was applied to obtain unique features to the embedding of the systems. We note that each of the embedded vectors corresponds to a systems. In this study, the dimension $n$ was chosen to be three.

\subsection{Simple ML Model to Predict Viscosity}
Each of the embedded vectors was regarded as features for the system, i.e. each system was represented by $n$ features. These features were regressed to the viscosity using the following equation:
\begin{equation}
    \eta_{\rm{pred}} = w_0 + w_1 \ln(x_1+\alpha) + \sum^{n}_{i=2} w_i x_i
    \label{eq:model}
\end{equation}
where $\eta_{pred}$ is the predicted viscosity, $x_i$ is the $i$th feature, $w_i$ is the wight of $x_i$, $w_0$ is the bias parameter and $\alpha$ is the bias parameter of $x_1$. We selected the logarithm scale for the first principal component (PC1) based on the observation that PC1 varied on the logarithmic scale equivalently to wall sliding velocity (see Result). The wight parameters were optimized using the following equation:
\begin{equation}
    w_0, ..., w_n = \argmin_{w_0, ..., w_n} \sum_{i=1}^{n}(\eta_{\rm{pred}} - \eta_{\rm{true}})^2
\end{equation}
where $w_i$ was determined using linear regression, and the bias $\alpha$ was determined to maximize the negative correlation between $\ln(x_1+\alpha)$ and $\eta_{\rm{true}}$. The calculated viscosity $\eta_{\rm{true}}$ was obtained from Kobayashi et al \cite{kobayashi2021}, which were calculated from the stress tensor of the sliding wall.

\subsection{Detection of Characteristic Behavior}
As Equation (\ref{eq:model}) is a function of the embedding vectors, an interpretation of the embedding vectors in terms of molecular movements is necessary to relate the latter to the resulting viscosity. For this, we introduce the function $g(\vect{x}_i)$ such that,
\begin{equation}
        g_{ij}({\vect{x}_i})= \mathbb{E}_{\vect{x}' \sim \vect{y_j} }\left[ f_{ij}^*(\vect{x}_i) - f_{ij}^*(\vect{x}')\right].\label{eq:gx}
\end{equation}
The function $g(\vect{x}_i)$ is the measures of characteristics for a specific molecular movement in system $i$, compared to the average dynamics in the other system $j$, i.e. it is a measurement of the contribution of $\vect{x}_i$ to $W_{ij}$ (Fig. 1d). Thus, $W_{ij}$ can equivalently be written as,   
\begin{equation}
    W_{ij}= \mathbb{E}_{\vect{x} \sim \vect{y_i} }\left[ g(\vect{x}_i)\right]. \label{w-gx}
\end{equation}
If the function $g(\vect{x}_i)$ specifies the characteristic dynamics that enlarge $W_{ij}$, then the characteristic molecular movement is also related to the PCs. For instance, if two systems are distant only along the PC1 in the embedding map, the difference in the PC1 reflects the Wasserstein distance. Hence the characteristic molecular movement contributes to the differences in PC1.    

\section{Result and Discussion}
\subsection{Embedding Map}
We characterized the molecular movements of the 16 different systems using the LDE, calculated the relative differences and obtained the embedded vectors using the ML methods. In this section, we present an embedding map that can differentiate the molecular movements of the 16 systems. Fig. \ref{fig:emb} illustrates the embedded vectors of 16 systems in the PC1--PC2 plane. For PC1, the systems with varying wall sliding speed located on the right side, which was true for all lubricant types. The logarithmic scale on the PC1 axis indicates that the molecular movements equally changed with the change of speed in the sliding wall. Conversely, PC2 of the alkane lubricants (hexadecane, PAO, pristane) and OMCTS diverged with an increase in sliding speed with each other. The alkane lubricants moved from the left -middle to the right-bottoms with the increase of the speed. In contrast, PC2 of OMCTS systems shifted the other way, i.e. from the middle-left to the top-right. This suggests that similar and contrasting properties of the alkane lubricants and OMCTS could be decoupled by PC1 and PC2, respectively. 

\subsection{Viscosity Prediction}
The PC1 trend agrees with the shear thinning effect, where the viscosity of a fluid decreases in response to an increase in the shear velocity. To quantitatively evaluate the relationship between the PC1 and viscosity, the correlation between the logarithm of PC1 and the logarithm of the viscosity was calculated, and was observed to be a strongly negative ($r=-0.94$; see Fig. \ref{fig:pc1_visc}). Furthermore, all three principal components were fitted into the ML model using Equation (\ref{eq:model}) to predict viscosity. The magnitude of correlation increased slightly to $r=0.95$ as shown in Fig. \ref{fig:vis_fit}. Considering that the inverse of the positive and negative sign between the two correlation coefficients is due to the weight parameter for PC1, the viscosity is presumably determined just by PC1. While the recent study showed that stress tensors from NEMD simulations can show the viscosity \cite{kadupitiya2021probing}, our results suggests that the viscosity is presumed similarly from the molecular movements.

\subsection{Interpretation of the Embedding Map}
To interpret the differences in the molecular movements indicated in the PC1 and PC2, we specified the characteristic molecular movements using $g(\vect{x})$. In this section, we compared low-viscosity systems to high-viscosity system. First, the system consisting of OMCTS and $V=1$ m/s was compared to the system with PAO and $V=0.001 $ m/s. The two systems have the largest PC1 difference. The probability distribution of $g(\vect{x})$ of the latter system was used as the reference because it had a more or less normal distribution. This implies that the molecular movement whose $g(\vect{x})$ is equal to the average $g(\vect{x})$ over the distribution is representative of the cases with slow sliding speeds. The probability distribution of $g(\vect{x})$ for OMCTS with $V=1 $ m/s has a skew to the right, indicating that the characteristic behavior of the OMCTS and high shear system is rare (Fig. \ref{fig:gx_7_8}a). Fig. \ref{fig:gx_7_8}b shows that the characteristic trajectories are related to the magnitude of displacement in the streamwise direction: Thus $g(\vect{x})$ corresponds to the magnitude of displacement. Interestingly, some of the characteristic molecular movement was in the opposite direction to the wall sliding. A comparison of the wall-normal position and $g(\vect{x})$ is shown in Fig. \ref{fig:gx_7_8}c, in order to investigate the effect of the structure induced by the confinement. The distribution of $g(\vect{x})$ is broadly symmetric to the center of the middle plane between the walls. Some plots on low $g(\vect{x})$ are between the two layers, suggesting molecular movements of low $g(\vect{x})$ include layer transitions. These results indicate that, irrespective of the flow or anti-flow directions, larger movements are the characteristic molecular movements of OMCTS with $V=0.001 $ m/s whereas the small movements are the non-characteristic molecular movements, some of which include transitions between the layers. 

Next, we compared the PAO with $V=1 $ m/s to the PAO with $V=0.001 $ m/s. PC1 of the PAO with $V=1 $ m/s is similar to that of the OMCTS with $V=1 $ m/s, whereas the PC2s are quite different. In the PAO with $V=1 $ m/s system, the probability distribution of $g(\vect{x})$ has two peaks with almost equal values (Fig. \ref{fig:gx_8_11}a), which is indicative of two types of molecular movements. The higher $g(\vect{x})$ mode corresponds to molecular movements in the same speed as the sliding wall ($V=1$ m/s), whereas the lower mode corresponded to the static molecules (Fig. \ref{fig:gx_8_11}b). Fig. \ref{fig:gx_8_11}c shows that the value of $g(\vect{x})$ is strongly related to the position of the molecules along the direction normal to the wall, where molecular movements in the top layer shows the high $g(\vect{x})$ values. Similar results were observed for other alkane lubricants (see Fig. S3). These results indicate that alkane lubricants are absorbed into the top and bottom walls when the sliding speed is high.  

From the above two comparisons, the difference in PC1 was estimated to reflect the overall diffusion, whereas the mechanisms of diffusion were related to PC2. We note that there is no theoretical support for relationships of PC1--diffusion and PC2--mechanism, and the interpretations of PC1 and PC2 can vary with changes of molecule types and number of systems. The relation of PC1 and diffusion suggests that in the confined systems viscosity is largely determined by diffusion, as described in the reciprocal relationship between viscosity and diffusion length by the Stokes-Einstein equation. The differences in PC2 suggest that the mechanisms for the shear thinning in OMCTS and alkane lubricants are different. The sliding wall affects the OMCTS through absorption and disassociation, which is knows as the stick-slip mechanism \cite{tanaka2003molecular,lei2011stick,xu2018squeezing}. The wall can drag the absorbed OMCTS until the ordered structure of OMCTS is unable to sustain the deformation. At the critical point, disassociation between the wall and molecules occurs to maintain the ordered structure of the OMCTS. The potential energy in the OMCTS molecules is converted into kinetic energy of OMCTS molecules which dissipates as heat through the collisions with each others and atoms in the walls. Such collisions can induce motion that is opposite to the sliding direction. In contrast, alkane lubricants may interact more strongly with the wall than with each other. Consequently, the upper layer moves with the wall, while the lower layer stays static, leading to the separation of the upper and lower layers. This result is regarded as the extreme case of plug slip that was observed the previous research for flexible linear molecules in thicker layers under very fast shear (20 m/s) \cite{ewen2017effect}. These results suggest that slip between either the lubricant layers or the wall--lubricant interface is the key to shear thinning, which agrees with previously reported experimental studies \cite{smith2019solidification,rosenhek2015question,thompson1995structure}. In addition, the two types of the slip mechanisms support the previous study that proposed the independence between velocity profile and friction \cite{porras2018independence}.

\section{Conclusions}
We have presented the combination of an unsupervised ML method that represents molecular movements with features in an embedding map, and constructed a simple supervised ML model to predict viscosity based on these features. The simple model was applied to lubricant molecules in confined systems, and the model was successfully used to predict their viscosity. Furthermore, the differences in the features were interpreted by detecting the characteristic molecular movements in low-viscosity systems. We have found that the viscosity-related feature corresponded to the large diffusion. The other feature distinguished the different molecular mechanics, i.e. stick-skip and adhesion to the wall.   

The presented analysis can be applied to various other systems to investigate the relationship between the transport properties and molecular movements. The influence of additives on lubricants can be related to the properties of the lubricant via the induced molecular movements of lubricants. This approach is also applicable to other materials where transport properties are crucial such as ions in batteries and gases molecules through selective membranes. We believe that ML-assisted analysis of molecular movements can extract important information. The movements would help bridge the gap between the static molecular structures and the transport properties, hence suggesting guidelines for material design.

\section*{Supporting Information}
Supporting information is available for free of charge. 
Streamwise displacement of UA particles and geometrical center for PAO and Pristane;
Distribution of $g(\vect{x})$ for PAO with $V$=1 m/s;
Distribution of $g(\vect{x})$, characteristic molecular movements and positions in wall-normal direction for PAO and Pristane with $V$=1 m/s.

\section*{Acknowledgments}
This work was supported in part by the Ministry of Education, Culture, Sports, Science and Technology (MEXT) as Research and Development of Next-Generation Fields.

\bibliography{apssamp}
 
\begin{figure}
\includegraphics[]{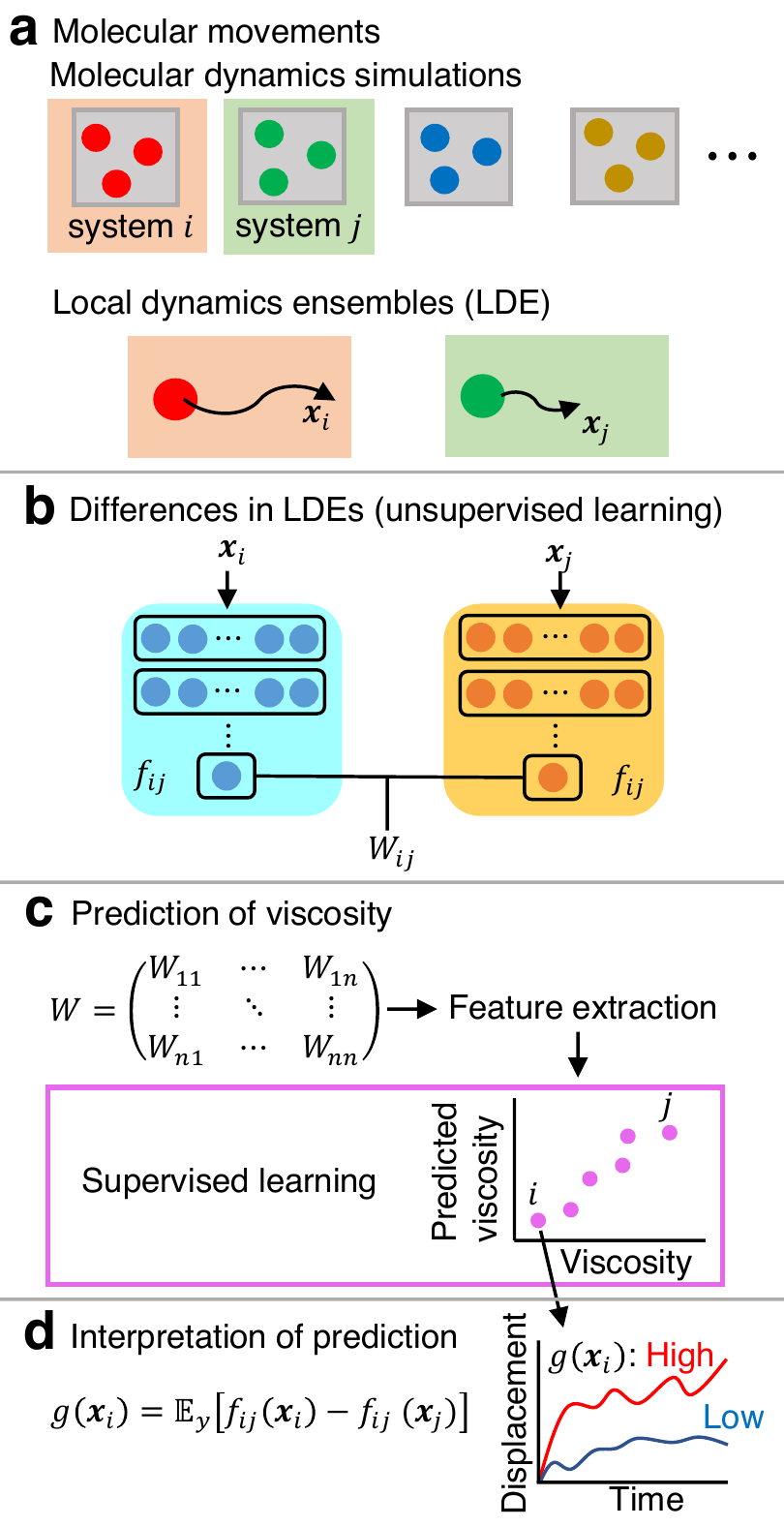}
\caption{ \label{fig:method} Scheme of machine learning (ML) to predict transport properties from molecular movements. (a) Molecular movements is obtained from molecular dynamics (MD) simulations, and represented by local dynamics ensemble (LDE). LDE is the ensemble of short-term trajectories $\vect{x}$ of the geometrical center of each molecule. (b) The differences of LDEs are calculated based on Wasserstein distance $W$ using deep neural networks $f$. (c) Wasserstein distances are calculated for all the pairs, and a distance matrix is obtained. Based on the distance matrix, features for each system are obtained, which are then used to predict viscosity. (d) The predicted viscosity is interpreted using $g(\vect{x})$, which can detect characteristic molecular movements that is related the features. If a molecular movement has a high $g(\vect{x}_i)$ value, it is characteristic to system $i$.}
\end{figure}

\begin{figure}
\includegraphics[]{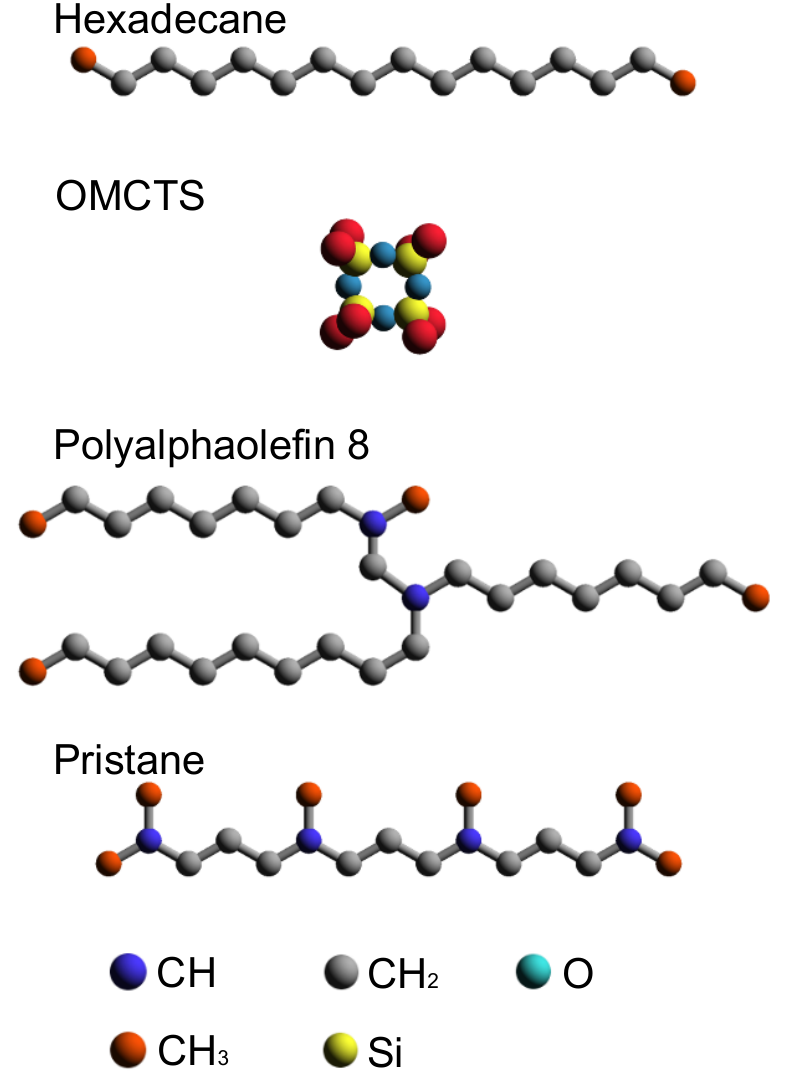}
\caption{ \label{fig:mols} United-atom (UA) models of the lubricant molecules, Hexadecane, Octamethylcyclotetrasiloxane (OMCTS), Polyalphaolefin 8 (PAO) and Pristane.}
\end{figure}

\begin{figure*}
\includegraphics[]{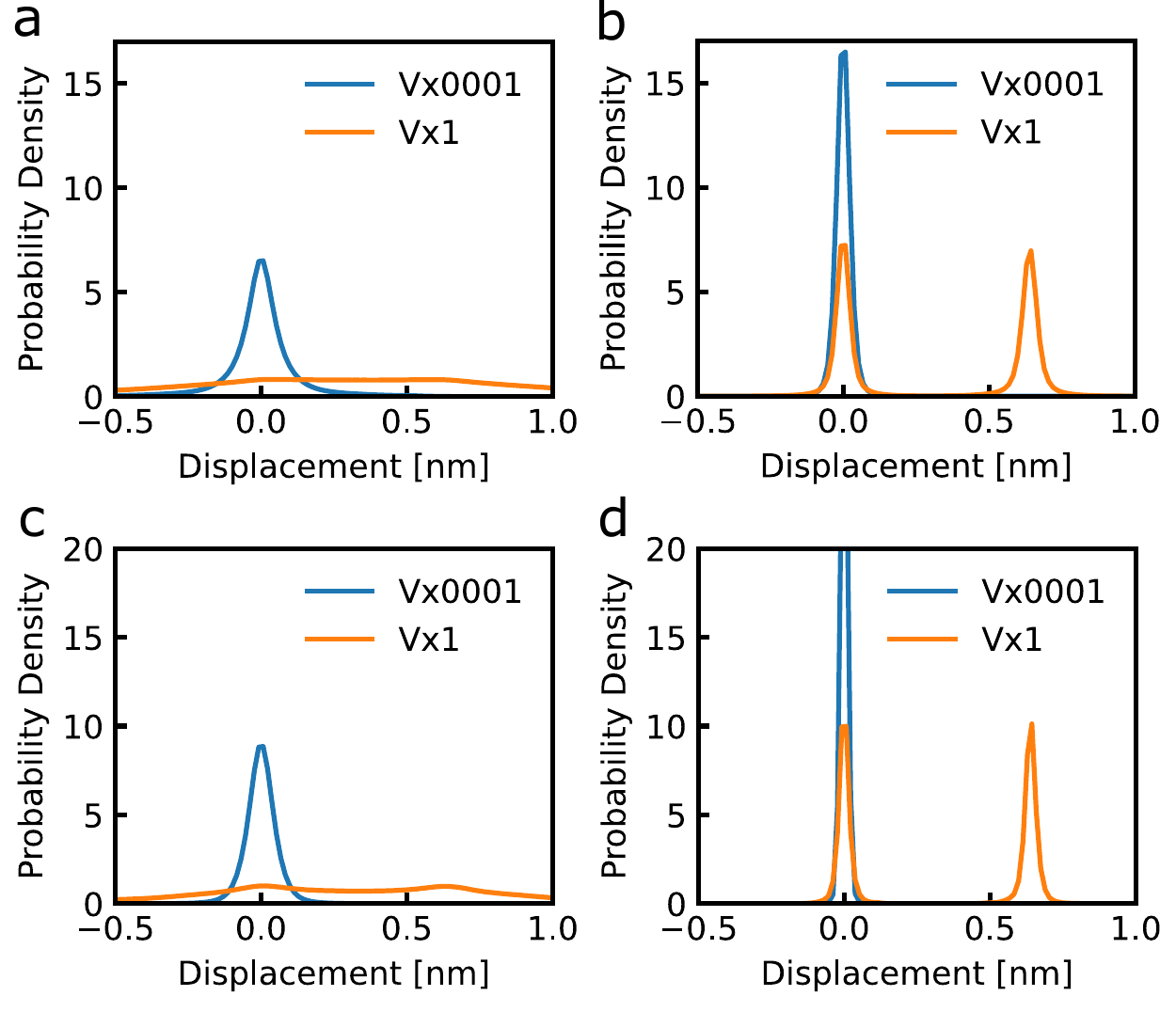}
\caption{ \label{fig:disp} Comparison of molecular movements between all particles in UA model and that of the geometrical center. The histogram shows the probability distributions of displacements in the streamwise direction at 640 ps. (a) UA particles of OMCTS. (b) UA particles of PAO. (c) The geometrical center of OMCTS. (d) The geometrical center of PAO.}
\end{figure*}

\begin{figure}
\includegraphics[]{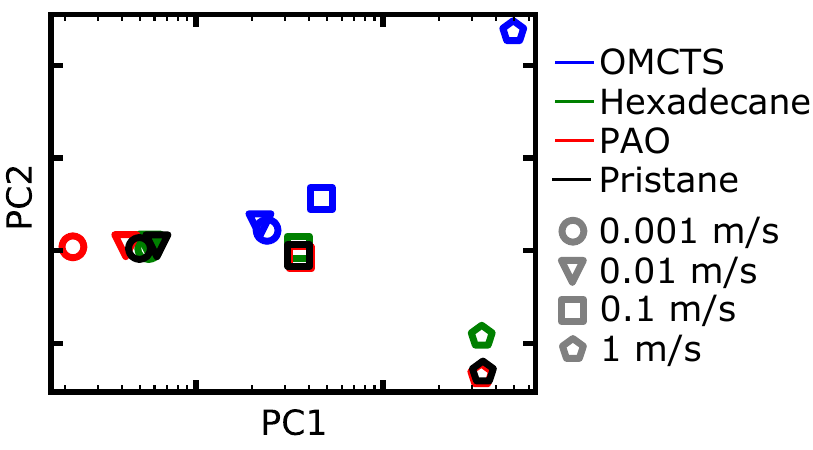}
\caption{ \label{fig:emb} Embedded vectors of the 16 systems. The logarithmic values of principal component (PC) 1 were obtained by adding a appropriate bias which was determined to minimize the correlation between PC1 and the calculated viscosity.}
\end{figure}

\begin{figure}
\includegraphics[]{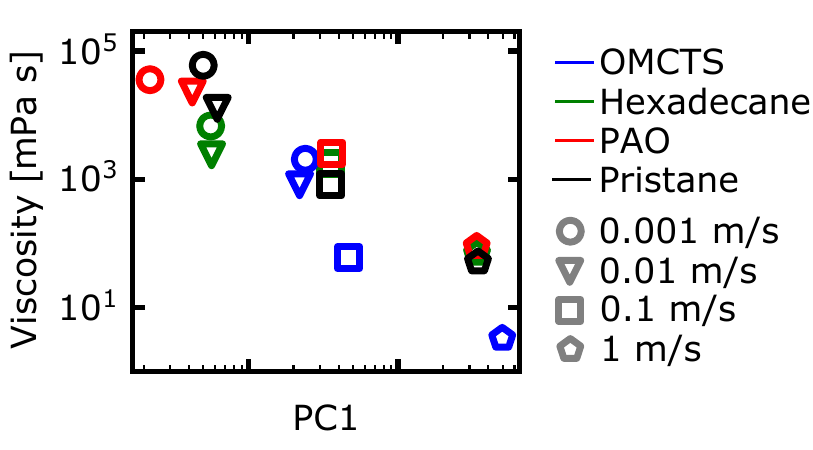}
\caption{ \label{fig:pc1_visc} The correlation between PC1 to viscosity. The viscosity was calculated by Kobayashi et al \cite{kobayashi2021} .}
\end{figure}

\begin{figure}
\includegraphics[]{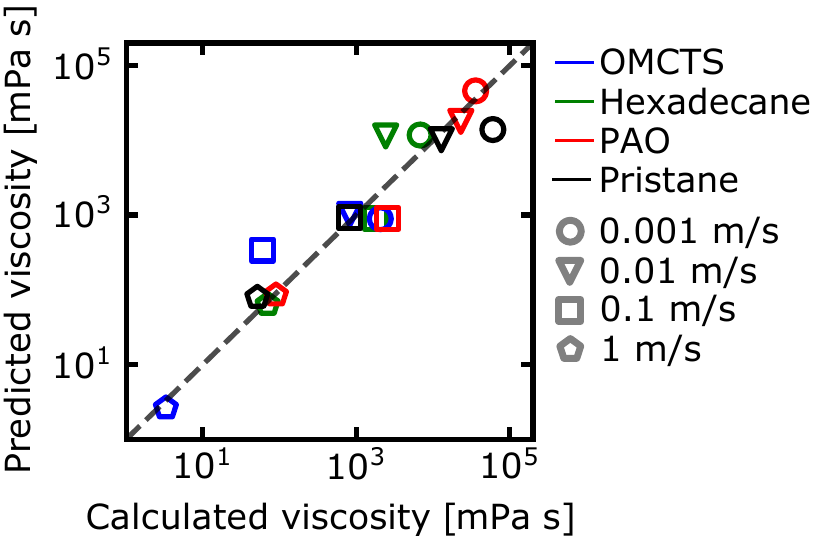}
\caption{ \label{fig:vis_fit} Prediction of viscosity using supervised ML model Equation (\ref{eq:model}).}
\end{figure}

\begin{figure}
\includegraphics{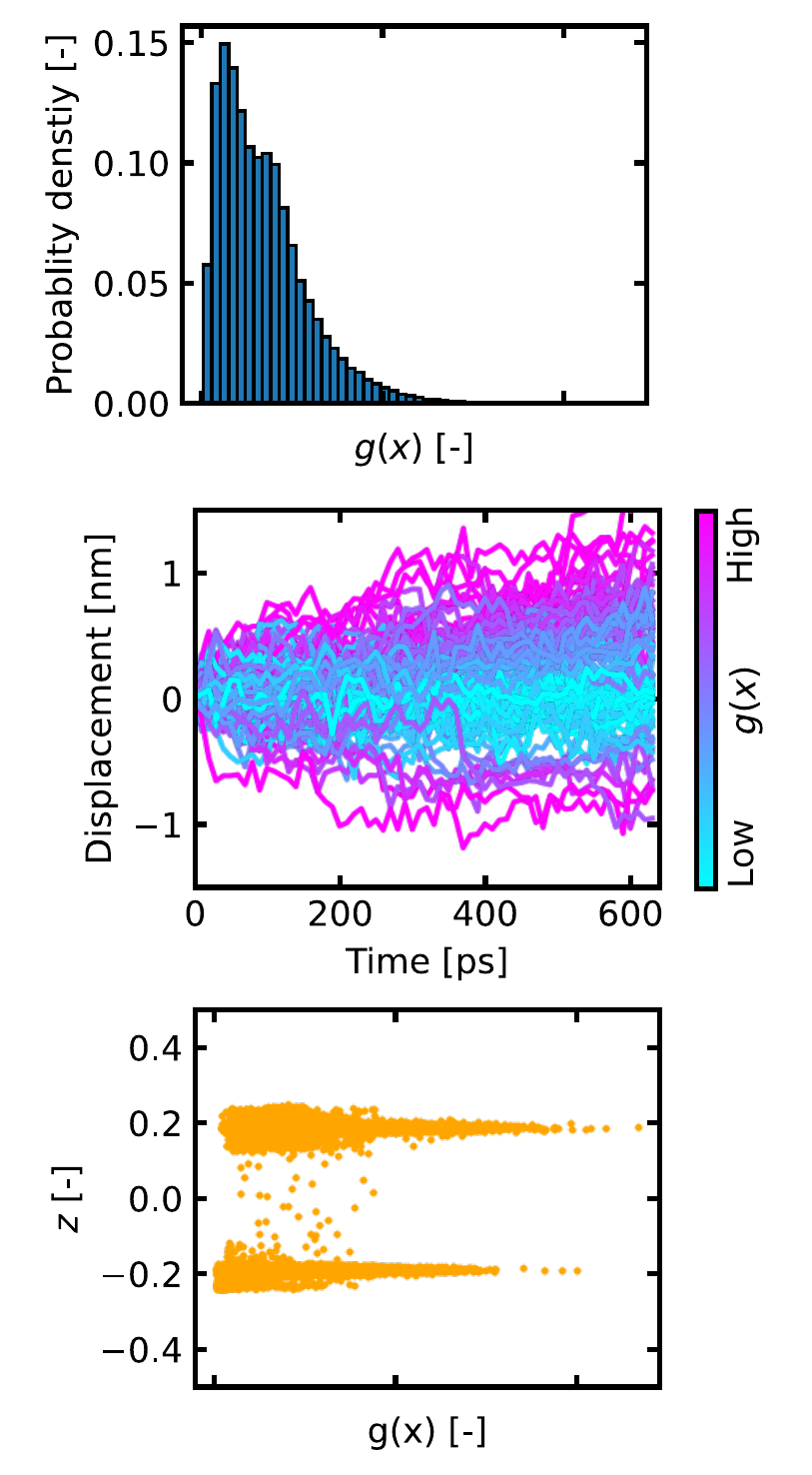}
\caption{ \label{fig:gx_7_8} Characteristic molecular movements of OMCTS with $V=1 $ m/s. The high viscosity system of PAO with $V=0.001 $ m/s was used for the reference system. (a) Histogram of the probability distribution of $g(\vect{x})$ in the system of OMCTS with $V=1 $ m/s. (b) History of displacements of the geometrical centers at 650 ps for 100 randomly selected trajectories. The trajectories are colored according to the value of $g(\vect{x})$. (c) Distribution of $g(\vect{x})$ along the position in normal direction normal to confined walls, i.e. $z$ axis. The position of $z$ axis was the average position for each short-term trajectory. The axis of $z$ was normalized by the wall thickness to make the range lie between -0.5 to 0.5.}
\end{figure}

\begin{figure}
\includegraphics[]{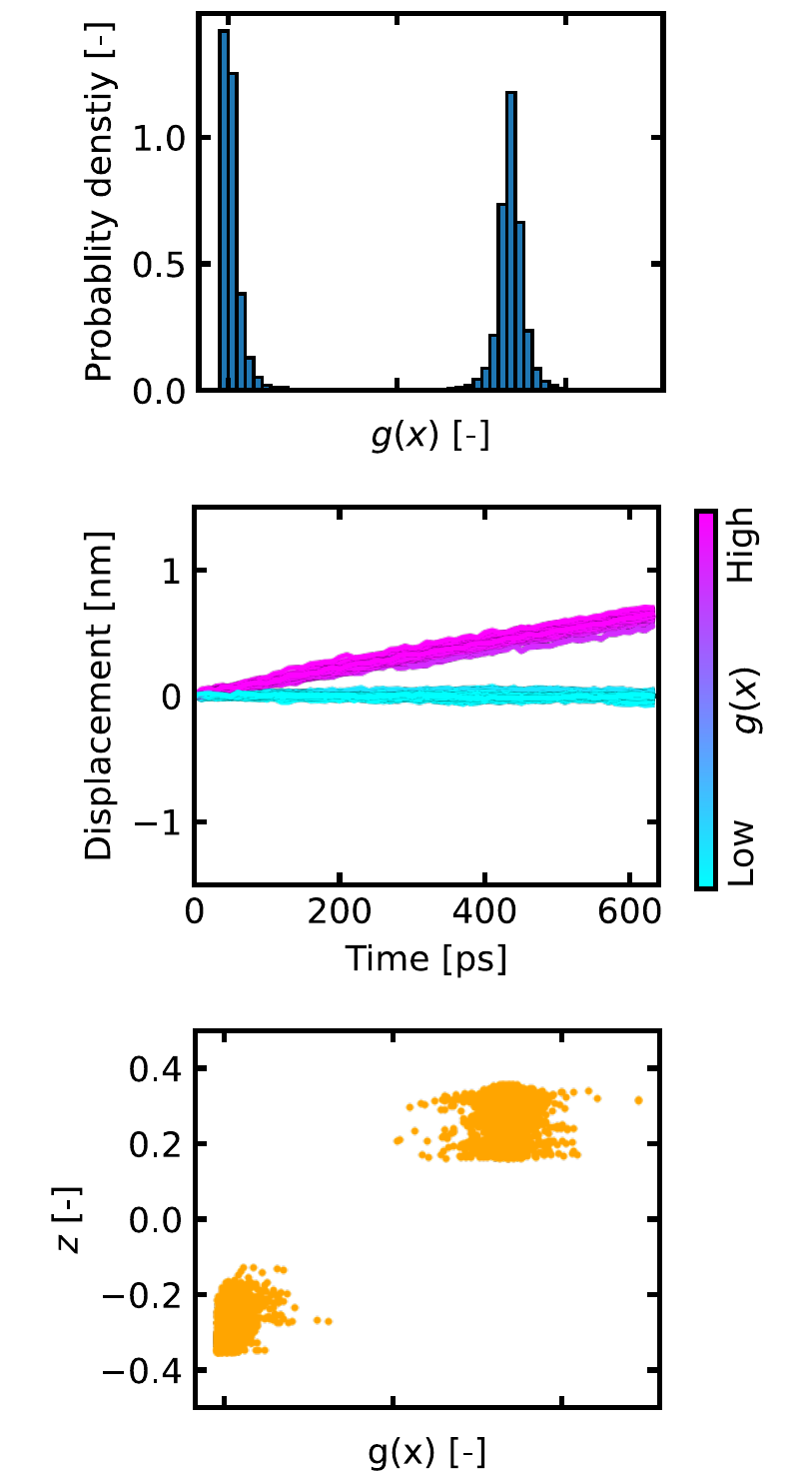}
\caption{ \label{fig:gx_8_11} Characteristic molecular movements of PAO with $V=1 $ m/s. The high viscosity system of PAO with $V=0.001 $ m/s was used for the reference system. (a) Histogram of the probability distribution of $g(\vect{x})$ in the system of PAO with $V=1 $ m/s. (b) History of displacements of the geometrical centers at 650 ps for 100 randomly selected trajectories. The trajectories are colored according to the value of $g(\vect{x})$. (c) Distribution of $g(\vect{x})$ along the position in normal direction normal to confined walls, i.e. $z$ axis. The position of $z$ axis was the average position for each short-term trajectory. The axis of $z$ was normalized by the wall thickness to make the range lie between -0.5 to 0.5.}
\end{figure}

\end{document}